\documentstyle [12pt]{article}
\def\p{\partial}
\def\d{\delta}
\def\l{\lambda}
\begin{document}
\thispagestyle{empty}
\begin{flushright}
 hep-th/9702144
\end{flushright}
\vspace{2cm}
\begin{center}
{\Large \bf On Infinite-Dimensional Algebras of Symmetries \\
\vspace{2mm}
of the Self-Dual Yang-Mills Equations}
\vspace{1cm}

{\large T.A.Ivanova}\footnote {e-mail: ita@thsun1.jinr.dubna.su}\\
\vspace{1cm}
{\em
Bogoliubov Laboratory of Theoretical Physics,\\
JINR, 141980 Dubna, Moscow Region, Russia
}\\
\end{center}
\vspace{1.5cm}
\begin{center}
{\large \bf Abstract}
\end{center}

\begin{quote}
Infinite-dimensional algebras of hidden symmetries of the
self-dual Yang-Mills equations are considered. A current-type
algebra of symmetries and an affine extension of conformal
symmetries introduced recently are discussed using the twistor
picture. It is shown that the extended conformal symmetries
of the self-dual Yang-Mills equations have a simple description
in terms of Ward's twistor construction.

\vspace{0.5cm}
\end{quote}
\newpage
{\bf I. INTRODUCTION}

\vspace{0.3cm}

The self-dual Yang-Mills (SDYM) equations in the space $R^4$
are manifestly invariant under the group of gauge transformations
and the group of conformal transformations of the space $R^4$.
Besides these symmetries, the SDYM equations have infinite-dimensional
algebras of `hidden symmetries'$^{1-6}$ related to gauge transformations.
In Refs.~1--5 the Yang gauge$^7$ in which two components of gauge
potential are equal to zero was used. Crane generalized the results of
Refs.~1--5 on the case when a gauge is not fixed and gave the
twistor interpretation of the action of symmetries on solutions of
the SDYM equations.$^8$

In the frames of approach of Refs.~1--4 it has been shown$^{9}$
that on the space $\cal M$ of local solutions of the SDYM equations
one can also define an (infinitesimal) action of the affine Lie algebra
related to conformal transformations of $R^4$. Later, the action of the
subalgebra of this algebra was also defined on the space of solutions
of the vacuum self-dual gravity equations.$^{10}$ In this paper we
shall show that the new symmetry algebra described in Ref.~9 has a simple
interpretation in terms of vector fields acting on transition matrices
of Ward's holomorphic vector bundles over twistor space.

\vspace{0.6cm}

{\bf II. DEFINITIONS AND NOTATION}

\vspace{0.3cm}

We consider the Euclidean space $R^{4,0}$ with the metric
$\d_{\mu\nu}$ and potentials  $A_\mu$ of the Yang-Mills (YM) fields
$F_{\mu\nu}=\p_\mu A_\nu - \p_\nu A_\mu + [A_\mu , A_\nu ]$, where
$\mu , \nu ,...=1,...,4$.  For simplicity we consider the fields
$A_\mu$ and $F_{\mu\nu}$ with values in the Lie algebra $su(n)$.

The SDYM equations have the form
$$
\frac{1}{2} \varepsilon _{\mu\nu\rho\sigma}F_{\rho\sigma}=F_{\mu\nu},
\eqno(1)
$$
where $\varepsilon _{\mu\nu\rho\sigma}$ is the completely
antisymmetric tensor on $R^{4,0}$ and $\varepsilon _{1234}=1$.

It is easily seen that eqs.(1) are invariant under the algebra of
infinitesimal local gauge transformations:
$$
\delta_\varphi A_\mu =\partial_\mu\varphi + [A_\mu , \varphi ],
\eqno(2)
$$
where $\varphi (x)\in su(n), \ x\in R^{4,0}$. It is also well-known
that the SDYM equations (1) are invariant with respect to the
algebra $so(5,1)$ of infinitesimal conformal transformations:
$$
\delta_N A_\mu =N^\nu\partial_\nu A_\mu + A_\nu\partial_\mu N^\nu,
\quad \partial_\mu :=\frac{\partial}{\partial x^\mu},
\eqno(3)
$$
where $N=N^\nu\partial_\nu$ is any generator of the 15-parameter
conformal group:
$$
X_a=\delta_{ab}\eta_{\mu\nu}^{b}x_\mu\partial_\nu ,\
Y_a=\delta_{ab}\bar\eta_{\mu\nu}^{b}x_\mu\partial_\nu ,\
P_\mu =\partial_\mu ,
$$
$$
K_\mu =\frac{1}{2}x_\sigma x_\sigma \partial_\mu -x_\mu D, \
D=x_\sigma\partial_\sigma,\ a,b,...=1,2,3,
\eqno(4)
$$
where $\{X_a\}$ and $\{Y_a\}$ generate two commuting $SO(3)$
subgroups in $SO(4)$, $P_\mu$ are the translation generators,
$K_\mu$ are the generators of special conformal transformations
and $D$ is the dilatation generator. Here $\eta_{\mu\nu}^{a}=
\{\varepsilon_{bc}^{a}, \mu =b, \nu =c;\ \delta_{\mu}^{a}, \nu =4;
\ -\delta_{\nu}^{a}, \mu =4\}$ is the self-dual 't Hooft tensor
and $\bar\eta_{\mu\nu}^{a}=
\{\varepsilon_{bc}^{a}, \mu =b, \nu =c;\ -\delta_{\mu}^{a}, \nu =4;
\ \delta_{\nu}^{a}, \mu =4\}$ is the anti-self-dual 't Hooft tensor.

Let $J=(J_{\mu}^{\nu})$  be the most general complex structure
on $R^{4,0}$:$^{11}$
$$
J_{\mu}^{\nu}= s_a\bar\eta_{\mu\sigma}^{a}\delta^{\sigma\nu}\
\Longrightarrow\ J_{\mu}^{\sigma}J_{\sigma}^{\nu}=-\delta_{\mu}^{\nu},
\eqno(5)
$$
where real numbers $s_a$ parametrize a two-sphere $S^2$, $s_as_a=1$.
Using $J$, one can introduce $(0,1)$ vector fields
$$
\bar V_{\bar 1}=\frac{1}{2}(\partial_1 + i\partial_2)-\frac{\lambda}{2}
(\partial_3 + i\partial_4)= \p_{\bar y}-\l\p_z,\
\bar V_{\bar 2}=\frac{1}{2}(\partial_3-i\partial_4)+\frac{\lambda}{2}
(\partial_1- i\partial_2)= \p_{\bar z}+\l\p_y,
\eqno(6)
$$
where $y=x_1+ix_2, z=x_3-ix_4, \bar y=x_1-ix_2, \bar z=x_3+ix_4$
are complex coordinates on $R^4\simeq C^2$, and $\lambda = (s_1+is_2)/
(1+s_3)$ is a local complex coordinate on $S^2\simeq CP^1$.

Let $S^1$ denote a contour $|\lambda |=1$ in the $\lambda$-plane
enclosing the origin in the counterclockwise direction, $C_+$
denote all complex numbers $\lambda\in CP^1$ with $|\lambda |\le
1+\alpha$, where $\alpha$ is any positive real number,
and $C_-$ denote all complex numbers $\lambda\in CP^1$ with $|\lambda |
\ge 1-\alpha$ (including $\lambda =\infty$).
Then $C_+$ and $C_-$ form a two-set open cover of the Riemann sphere
$CP^1$ with the intersection $C_\alpha = C_+\cap C_- = \{\lambda :\
1-\alpha \le |\lambda |\le 1+\alpha\}$. Vector field $\p_{\bar\lambda}:=
\p /\p_{\bar\l}$ is antiholomorphic $(0,1)$ vector field with respect to
the
standard complex structure $\varepsilon = i\;d\lambda\otimes \p_\lambda
-i\;d{\bar\lambda}\otimes \p_{\bar\lambda}$ on $S^2$.

Twistor space $\cal Z$ for $R^{4,0}$ is the space ${\cal Z}= R^{4,0}
\times CP^1\simeq CP^3 -CP^1 \subset CP^3$ which is the space of all the
constant complex structures on $R^{4,0}$.$^{12}$ This space can be covered
by two coordinate patches ${\cal Z}= U_+\cup U_-$:
$$
U_+=\{x\in R^{4,0}, \lambda\in C_+\},
\quad
U_-=\{x\in R^{4,0}, \lambda\in C_-\},
\eqno(7)
$$
with the intersection $U= U_+\cap U_-=\{x\in R^{4,0}, \lambda\in
C_\alpha =C_+\cap C_-\}$.

Twistor space $\cal Z$ is a trivial bundle $\pi :\ {\cal Z}\rightarrow
R^{4,0}$ over $R^{4,0}$ ($\pi :\ \{x_\mu ,\l ,\bar\lambda\}\rightarrow
\{x_\mu\}$) with fibre $CP^1$ and is a complex manifold with complex
structure ${\cal J}=(J,\varepsilon )$ on $\cal Z$. Vector fields
$\bar V_{\bar 1}, \bar V_{\bar 2}$ from (6) and $\bar V_{\bar 3}=
\p_{\bar\l}$ are vector fields of type $(0,1)$ with respect to complex
structure $\cal J$. Holomorphic coordinates on $U_+$ and $U_-$ are
$\{v_{+}^{1}=y-\l\bar z, v_{+}^{2}=z+\l\bar y, v_{+}^{3}=\l \}$
and $\{v_{-}^{1}=\frac{1}{\l}y-\bar z, v_{-}^{2}=\frac{1}{\l}z+\bar y,
v_{-}^{3}=\frac{1}{\l}\}$. Twistor space is also a nontrivial
holomorphic bundle $p:\ {\cal Z}\rightarrow CP^1$ over $CP^1$
with holomorphic sections (projective lines $CP_{x}^{1}$)
parametrized by points $x\in C^4$:
$$
CP_{x}^{1}=\{ v_+^1=y+\l\tilde z,\ v_+^2=z+\l\tilde y,\ (y,z,\tilde y,
\tilde z)\in C^4\}.
\eqno(8)
$$
On $\cal Z$ one can introduce a real structure as a map $\tau :\
R^{4,0}\times S^2 \rightarrow R^{4,0}\times S^2$ defined by the formula:
$\tau (x_\mu ,\l )=(x_\mu , -1/{\bar\l}),\ x_\mu\in R^{4,0}$.
Then, for real holomorphic sections of the bundle ${\cal Z}\rightarrow
CP^1$ satisfying the equations $\tau (v_{+}^{1})=\overline{v_{-}^{2}},\
\tau (v_{+}^{2})=-\overline{v_{-}^{1}}$ we have $\tilde z= -\bar z,\
\tilde y= \bar y$.$^{12, 11}$

\vspace{0.6cm}

{\bf III. WARD'S TWISTOR CORRESPONDENCE AND }

{\bf $~~~~~~$GAUGE-TYPE HIDDEN SYMMETRIES}

\vspace{0.3cm}

The potentials $A_\mu$ define a connection $D:=dx^\mu (\p_\mu +A_\mu )$
in the principal $SU(n)$-bundle $P=P(R^{4,0}, SU(n))$ over $R^{4,0}$.
In the standard manner we introduce the complex vector bundle
$E=P\times_{SU(n)}C^n$ associated to $P$. Sections of this bundle are
$C^n$-valued vector-functions depending on $x\in R^{4,0}$. Using
the projection $\pi :\ {\cal Z}\rightarrow R^{4,0}$, we can pull back
the bundle $E$ with self-dual connection $D$ to the bundle
$\tilde E:=\pi^*E$ over $\cal Z$, and the pulled back connection
$\tilde{D}:=\pi^*D$ will be flat along the fibres $CP_{x}^{1}$
of the bundle ${\cal Z}\rightarrow R^{4,0}$. We can take it in the form
$\tilde{D}= {D}+d\l\p_\l + d\bar\l\p_{\bar\l}$.

Let us consider local holomorphic sections $\xi$ of the bundle
$\tilde E$ or, in other words, local solutions of the equations
$\tilde{D}_{\bar a}^{(0,1)}\xi =0$:
$$
[(D_1 + iD_2)-\l (D_3 +iD_4)]\xi (x,\l ,\bar\l )=0,
\eqno(9a)
$$
$$
[(D_3 - iD_4)+\l (D_1 -iD_2)]\xi (x,\l ,\bar\l )=0,
\eqno(9b)
$$
$$
\p_{\bar\l}\xi (x,\l ,\bar\l )=0,
\eqno(10)
$$
where $\tilde{D}_{\bar a}^{(0,1)}$ are components of $\tilde D$
along $(0,1)$ vector fields $\bar V_{\bar a}$. We can solve
eq.(10), and then eqs.(9) on $\xi (x,\lambda )$ are usually called the
linear system for the SDYM equations.$^{13, 14}$ It is easily seen that
the compatibility  conditions of the linear system (9) coincide with the
SDYM equations (1).

Equations (9) have a local solution $\xi_+(x,\l )$ over
$U_+\subset\cal Z$, a local solution $\xi_-(x,\l )$ over
$U_-\subset\cal Z$ and $\xi_+=\xi_-$ on $U_+\cap U_-$ . We can always
represent $\xi_\pm$ in the form $\xi_\pm =\psi_\pm\chi_\pm$, where
$SL(n, C)$-valued functions $\psi_\pm$ are holomorphic in $\l$ on
$U_\pm$, and vector-functions $\chi_\pm\in C^n$ defined on $U_\pm$
are related by
$$ \chi_-={\cal F}\chi_+ \eqno(11) $$
on $U=U_+\cap U_-\subset \cal Z$. Here $\chi_+=\chi_+(v_{+}^{a})$
and $\chi_-=\chi_-(v_{-}^{a})$ are \v{C}ech fibre coordinates of
the bundle $\tilde E$ over $U_+$ and $U_-$, and $\cal F$ is the
transition matrix in the bundle $\tilde E$. Matrix $\cal F$ is a
holomorphic $SL(n, C)$-valued function on $U$ with non-vanishing
determinant.  By Ward's construction, the bundle $\tilde E$ is
holomorphically trivial on each fibre $CP_{x}^{1}$, i.e. $\tilde
E\!\mid_{CP_{x}^{1}}=CP_{x}^{1}\times C^n$.$^{14}$ This means that
on $CP_{x}^{1}$ matrix $\cal F$ can be factorized in the form
$$
{\cal F}=\psi^{-1}_-(x,\l )\psi_+(x,\l )
\eqno(12)
$$
for each point $x$ from an open subset of $R^{4,0}$.

{\it Remark}. Let we are given a matrix-valued function ${\cal F}\in
G$ on $C_\alpha =C_+\cap C_-\subset CP^1$. Then Riemann-Hilbert
problem is to find matrix-valued functions $\psi_+\in G$ and
$\psi_-\in G$ such that $\psi_+$ is regular (i.e. holomorphic with
non-vanishing determinant) on $C_+$, $\psi_-$ is regular on $C_-$ and
${\cal F}=\psi^{-1}_-\psi_+$ is regular on $C_\alpha = C_+\cap C_-$.
Solution of the Riemann-Hilbert problem exists (Birkhoff's theorem)
and is unique up to the left multiplication by matrix $g$ not
depending on $\l :\ \psi_\pm \rightarrow g\psi_\pm $ (for discussion
see, e.g., Ref.~15).  Thus, the splitting  (12) gives a solution of a
parametric Riemann-Hilbert problem ($x_\mu$ are external parameters).
Besides, the matrix $\cal F$ in (12) will not change if we replace
$\psi_\pm (x,\l )$ by $g(x)\psi_\pm (x,\l )$ (standard gauge
transformation).

Matrix-valued functions $\psi_\pm $ define a trivialization of the
bundle $\tilde E$ over $U_\pm$. It follows from (9)--(12) that
$$
(\p_{\bar y}\psi_+ - \l \p_z\psi_+)\psi_+^{-1}=
(\p_{\bar y}\psi_- - \l \p_z\psi_-)\psi_-^{-1}=
-(A_{\bar y}(x) - \l A_z(x)),
\eqno(13a)
$$
$$
(\p_{\bar z}\psi_+ + \l \p_y\psi_+)\psi_+^{-1}=
(\p_{\bar z}\psi_- + \l \p_y\psi_-)\psi_-^{-1}=
-(A_{\bar z}(x) + \l A_y(x)),
\eqno(13b)
$$
and the potentials $\{A_\mu\}$ defined by (13) satisfy the SDYM equations.

{}For a Hermitian vector bundle when the structure group is $SU(n)$
the real structure $\tau$ on the twistor space $\cal Z$ induces a
Hermitian structure in the bundle $\tilde E$.$^{12}$ Then, fields
$\{A_\mu\}$, matrices $\psi_\pm$ and $\cal F$ have to satisfy the
following unitarity conditions (see e.g. Ref.~8):
$$
A_{y}^{\dagger }=-A_{\bar y},\ A_{z}^{\dagger }=-A_{\bar z},\
A_{\bar y}^{\dagger }=-A_{y},\ A_{\bar z}^{\dagger }=-A_{z},
\eqno(14a)
$$ $$
\psi_{+}^{\dagger }(\l )=\psi_{-}^{-1}(-\frac{1}{\bar\l}),\
\psi_{-}^{\dagger }(\l )=\psi_{+}^{-1}(-\frac{1}{\bar\l}),
\eqno(14b)
$$ $$
{\cal F}^\dagger (\l )={\cal F}(-\frac{1}{\bar\l}),
\eqno(14c)
$$
where $^\dagger $ denotes Hermitian conjugation.

{}From (13) one can see that gauge potentials $A_\mu$ do not change
after transformations
$$
\psi_+\rightarrow\psi_{+}^{eqv}=\psi_+h_+,\
\psi_-\rightarrow\psi_{-}^{eqv}=\psi_-h_-,
\eqno(15)
$$
where $h_+$ is any regular holomorphic matrix-valued function on $U_+$,
and $h_-$ is any regular holomorphic matrix-valued function on $U_-$.
This means that the bundles with transition matrices $h_{-}^{-1}{\cal
F}h_+$ and $\cal F$ are holomorphically equivalent. Notice that for
the Hermitian vector bundles $\tilde E$ with transition matrices
$\cal F$ satisfying the `hermiticity' condition (14c) the equality
$h_{+}^{\dagger }(\l )= h_{-}^{-1}(-1/{\bar\l})$ has to be
fulfilled.

Thus, we have described a one-to-one correspondence between gauge
equivalence classes of solutions to the SDYM equations on the
Euclidean 4-space and equivalence classes of holomorphic vector
bundles $\tilde E$ over twistor space $\cal Z$, that are
holomorphically trivial over each real projective line $CP_{x}^{1}$
in $\cal Z$ (the Euclidean version of Ward's theorem$^{16, 12, 11}$).

Now we will briefly remind the description of gauge-type `hidden
symmetries'. First, we should define an infinite-dimensional complex
Lie (pseudo)group $H$ (current-type
group) of holomorphic maps from  $U=U_+\cap U_-\subset \cal Z$ to the
group $SL(n, C)$ with standard pointwise multiplication and the
corresponding complex Lie algebra $\cal H$ of holomorphic maps from
$U$ to the Lie algebra $sl(n, C)$. We shall also consider subgroups
$H_\pm\subset H$ of elements from $H$ which can be extended
continuously to holomorphic maps from $U_\pm$ to $SL(n, C)$ and
the corresponding algebras ${\cal H}_\pm\subset \cal H$.

The action of group $H$ on transition matrix $\cal F$ preserving the
hermiticity condition (14c) is given by:$^{8}$
$$
{\cal F}\rightarrow \tilde{\cal F}=u(h){\cal F}:=h(\l ){\cal F}(\l )
h^\dagger (-\frac{1}{\bar\l}),
\eqno(16)
$$
where
$$h(\l )\equiv h(y-\l\bar z, z+\l\bar y, \l ), {\cal F}\equiv
{\cal F}(y-\l\bar z, z+\l\bar y, \l ),  h^\dagger (-\frac{1}{\bar\l})
\equiv h^\dagger (y+\frac{1}{\bar\l}\bar z, z-\frac{1}{\bar\l}\bar y,
-\frac{1}{\bar\l}),$$
 and in (16) we simply omitted a part of arguments.
To this action the following action of the Lie algebra $\cal H$ on
$\cal F$ corresponds:
$$
{\cal F}\rightarrow \delta_\varphi {\cal F}
=\varphi (\l ) {\cal F}(\l ) +{\cal F}(\l )\varphi^\dagger
(-\frac{1}{\bar\l}),
\eqno(17) $$
where $\varphi (\l )\equiv \varphi (y-\l\bar z, z+\l\bar y, \l ) \in
sl(n,C)$ is any element of $\cal H$ (of course, $\varphi^\dagger
(-\frac{1}{\bar\l})$ is also an element of $\cal H$).

Consider now the following $sl(n,C)$-valued function $\phi$ on $U$
$$
\phi\equiv \varphi_--\varphi_+:=\psi_-(\delta_\varphi{\cal F})
\psi_{+}^{-1}=\psi_-\varphi (\l )\psi_{-}^{-1}+ \psi_+\varphi^\dagger
(-\frac{1}{\bar\l})\psi_{+}^{-1},
\eqno(18)
$$
where for holomorphic $\phi$ expanded in Laurent series $\phi =
\sum_{n=-\infty}^{\infty}\l^n\phi_n(x)$ we put
$$
\varphi_+:=\tilde\phi_{0}(x)-\sum_{n=1}^{\infty}\l^n\phi_n(x),\
\varphi_-:=\hat\phi_{0}(x)+\sum_{n=-\infty}^{-1}\l^n\phi_n(x),\
\hat\phi_{0}(x)-\tilde\phi_{0}(x)=\phi_{0}(x).
\eqno(19)
$$
Thus, the function $\varphi_+(\l )\in sl(n,C)$ is holomorphic in
$\l\in C_+\subset CP^1$, and the function $\varphi_-(\l )\in sl(n,C)$
is holomorphic in $\l\in C_-\subset CP^1$.

It follows from (13) and (18) that
$$
(D_{\bar y}-\l D_z)\phi \equiv (\p_{\bar y}-\l \p_z)\phi +
[A_{\bar y} -\l A_z, \phi ]=0\ \Longrightarrow\
(D_{\bar y}-\l D_z)\varphi_+=(D_{\bar y}-\l D_z)\varphi_-,
\eqno(20a)
$$
$$
(D_{\bar z}+\l D_y)\phi \equiv (\p_{\bar z}+\l \p_y)\phi +
[A_{\bar z} +\l A_y, \phi ]=0\ \Longrightarrow\
(D_{\bar z}+\l D_y)\varphi_+=(D_{\bar z}+\l D_y)\varphi_-.
\eqno(20b)
$$
The action of algebra $\cal H$ on matrix-valued functions $\psi_\pm
\in SL(n,C)$ and on gauge potentials $\{A_\mu\}$ is given by formulae
$$
\delta_\varphi\psi_+:=-\varphi_+\psi_+,\
\delta_\varphi\psi_-:=-\varphi_-\psi_- ,
\eqno(21)
$$
$$
\delta_\varphi A_{\bar y}-\l\delta_\varphi A_z:=
D_{\bar y}\varphi_+-\l D_z\varphi_+
=D_{\bar y}\varphi_--\l D_z\varphi_- ,
\eqno(22a)
$$
$$
\delta_\varphi A_{\bar z}+\l\delta_\varphi A_y:=
D_{\bar z}\varphi_++\l D_y\varphi_+
=D_{\bar z}\varphi_-+\l D_y\varphi_- .
\eqno(22b)
$$
It follows from (22) that
$$
\delta_\varphi A_y=\oint_{S^1}\frac{d\l}{2\pi i\l^2}
(D_{\bar z}\varphi_+ +\l D_y\varphi_+ ) ,\
\delta_\varphi A_z=-\oint_{S^1}\frac{d\l}{2\pi i\l^2}
(D_{\bar y}\varphi_+ -\l D_z\varphi_+ ) ,
$$ $$
\delta_\varphi A_{\bar y}=\oint_{S^1}\frac{d\l}{2\pi i\l}
(D_{\bar y}\varphi_+ -\l D_z\varphi_+ ) ,\
\delta_\varphi A_{\bar z}=\oint_{S^1}\frac{d\l}{2\pi i\l}
(D_{\bar z}\varphi_+ +\l D_y\varphi_+ ),
\eqno(23)
$$
where $S^1=\{\l\in CP^1 : |\l |=1\}$. Thus, in (17), (21) and (22) we
have described the action of the current algebra $\cal H$ on the space
of local solutions of the SDYM equations.

\vspace{0.6cm}

{\bf IV. HIDDEN SYMMETRIES FROM THE AFFINIZATION}

{\bf $~~~~~$OF CONFORMAL ALGEBRA }

\vspace{0.3cm}

In Ref.~9 to each generator $N$ of the group $SO(5,1)$ of conformal
transformations of the Euclidean space $R^{4,0}$ Popov and
Preitschopf associated an infinite number of infinitesimal symmetry
transformations $\delta_{N}^{n}\ (n=0,1,2,...)$ acting on solutions
of the SDYM equations.  They considered potentials $A_\mu$ with
values in complex Lie algebras and did not discuss the problems of
reality for gauge fields.  Considering potentials $A_\mu$ with values
in $su(n)$, we examine not only the infinitesimal symmetries from
Ref.~9, which do not preserve, generally speaking, the unitarity
conditions (14), but also the symmetries preserving the unitarity.

In formulae (4) we have specified a homomorphism of the Lie algebra
$so(5,1)$ of the conformal group into the Lie algebra of vector fields
$M,N,...$ on $R^{4,0}$. Now we have to define the action of $SO(5,1)$
on $\cal Z$, which preserves the holomorphicity of the bundle
$\tilde E \rightarrow \cal Z$. Such lift $N\rightarrow\tilde N$ of
vector fields on $\cal Z$ was described in Ref.~17, and the lifted
generators
$$
\tilde X_a=X_a,\ \tilde Y_a=Y_a+2Z_a,\ \tilde P_\mu=P_\mu ,\
\tilde K_\mu=K_\mu+\bar\eta_{\sigma\mu}^{a}x_\sigma Z_a,\
\tilde D=D,
\eqno(24)
$$
$$
Z_1=\frac{1}{2} i(\l^2-1)\p_\l -\frac{1}{2} i(\bar\l^2-1)\p_{\bar\l},\
Z_2=\frac{1}{2}(\l^2+1)\p_\l +\frac{1}{2}(\bar\l^2+1)\p_{\bar\l},\
Z_3=i\l\p_\l-i\bar\l\p_{\bar\l}
\eqno(25)
$$
are infinitesimal automorphisms of the complex structure $\cal J$
on $\cal Z$. This means, in particular, that for any generator
$\tilde N$ from (24) the following relations
$$
[\bar V_{\bar a},\tilde N]=\alpha_{\bar a}^{\bar b}(\tilde N)
\bar V_{\bar b}
\eqno(26) $$
take place for some functions $\alpha_{\bar a}^{\bar b}(\tilde N) $
and vector fields $\bar V_{\bar 1}, \bar V_{\bar 2}$ from (6) and
$\bar V_{\bar 3}=\partial_{\bar\lambda}$.

Having generators $\tilde N$ of infinitesimal holomorphic diffeomorphisms
(24) of the space $\cal Z$, we consider the holomorphic vector fields
$N_n=\l^{-n}\tilde N,\ n=0, \pm 1, \pm 2,...$, and define the following
infinitesimal transformations of transition matrix $\cal F$ of the bundle
$\tilde E$
$$
\delta_{N}^{n}{\cal F}:=N_n({\cal F})=\l^{-n}\tilde N({\cal F})
,\ n=0, \pm 1, \pm 2,...\  .
\eqno(27)
$$
In virtue of (26) transformations (27) are holomorphic and, therefore,
define the infinitesimal symmetries of the SDYM equations.

Let us introduce the $sl(n, C)$-valued function (compare with (18) from
Sec. III)
$$
\phi_{N_n}=\phi_{N_n}^--\phi_{N_n}^+:=\psi_-(\delta_{N}^{n}{\cal F})
\psi_{+}^{-1}=\l^{-n}(\tilde N\psi_+)\psi_{+}^{-1} -
\l^{-n}(\tilde N\psi_-)\psi_{-}^{-1},
\eqno(28)
$$
$$
\phi_{N_n}^+:=\tilde\phi_{N_n}^0(x)-\sum_{k=1}^{\infty}\l^k
\phi_{N_n}^k(x),
\ \phi_{N_n}^-:=\hat\phi^0_{N_n}(x)+
\sum_{k=-\infty}^{-1}\l^k\phi_{N_n}^k(x),\
\hat\phi^0_{N_n}- \tilde\phi_{N_n}^0=\phi_{N_n}^0.
\eqno(29)
$$
Remind that the splitting of the $sl(n,C)$-valued function
$\phi_{N_n}(x,\l )$ into the difference of the function
$\phi_{N_n}^{-}\in sl(n,C)$ holomorphic in $\l\in C_-$ and of the
function $\phi_{N_n}^{+}\in sl(n,C)$ holomorphic in $\l\in C_+$ is a
solution of the infinitesimal variant of Riemann-Hilbert problem.

{\it Remark}. Notice that
$\phi_{N_0}^{\pm}=-(\tilde N\psi_\pm)\psi_{\pm}^{-1}$,
but $\phi_{N_n}^{+}\ne -\l^{-n}(\tilde N\psi_+)\psi_{+}^{-1}$,
and $\phi_{N_n}^{-}\ne -\l^{-n}(\tilde N\psi_-)\psi_{-}^{-1}$
when $n\ne 0$. If we fix the gauge $\psi_+(\l =0)=1$, then the functions
$\phi_{N_0}^{+}$ coincide with the functions $-\psi_{\tilde N},
\tilde N\in so(5,1),$ which are generating functions for
symmetries introduced in Ref.~9.

It follows from (13), (26) and (28) that (cf. Sec. III):
$$
(D_{\bar y}-\l D_z)\phi_{N_n} =0
\ \Longrightarrow\
(D_{\bar y}-\l D_z)\phi^+_{N_n}=(D_{\bar y}-\l D_z)\phi^-_{N_n},
\eqno(30a)
$$
$$
(D_{\bar z}+\l D_y)\phi_{N_n} =0
\ \Longrightarrow\
(D_{\bar z}+\l D_y)\phi^+_{N_n}=(D_{\bar z}+\l D_y)\phi^-_{N_n},
\eqno(30b)
$$
$$
\delta_{N}^{n}\psi_+=-\phi_{N_n}^{+}\psi_+,\
\delta_{N}^{n}\psi_-=-\phi_{N_n}^{-}\psi_-,
\eqno (31)
$$
$$
\delta_{N}^{n}A_{\bar y}-\lambda\delta_{N}^{n}A_z:=
 D_{\bar y}\phi^+_{N_n}-\l D_z\phi^+_{N_n}=D_{\bar y}\phi^-_{N_n}
-\l D_z\phi^-_{N_n},
$$
$$
\delta_{N}^{n}A_{\bar z}+\lambda\delta_{N}^{n}A_y:=
 D_{\bar z}\phi^+_{N_n}+\l D_y\phi^+_{N_n}=D_{\bar z}\phi^-_{N_n}
+\l D_y\phi^-_{N_n}.
\eqno(32)
$$
We have
$$
\delta_{N}^{n} A_y=\oint_{S^1}\frac{d\l}{2\pi i\l^2}
(D_{\bar z}\phi^+_{N_n} +\l D_y\phi^+_{N_n}) ,\
\delta_{N}^{n} A_z=-\oint_{S^1}\frac{d\l}{2\pi i\l^2}
(D_{\bar y}\phi^+_{N_n} -\l D_z\phi^+_{N_n} ) ,
$$ $$
\delta_{N}^{n} A_{\bar y}=\oint_{S^1}\frac{d\l}{2\pi i\l}
(D_{\bar y}\phi^+_{N_n} -\l D_z\phi^+_{N_n} ) ,\
\delta_{N}^{n} A_{\bar z}=\oint_{S^1}\frac{d\l}{2\pi i\l}
(D_{\bar z}\phi^+_{N_n} +\l D_y\phi^+_{N_n} ),
\eqno(33)
$$
Thus, to each generator $N$ of the group of conformal transformations
of the space $R^{4,0}$ we have corresponded an infinite number of
generators $\delta_{N}^{n}$ of symmetry transformations of the SDYM
equations and defined their action on the transition matrix $\cal F$,
on the group-valued functions $\psi_\pm$ and on gauge potentials $A_\mu$.

By direct calculations, it is not hard to show that
$$
\delta_{N}^{0}A_\mu ={\cal L}_NA_\mu\equiv N^\nu\p_\nu A_\mu +
A_\nu\p_\mu N^\nu ,
\eqno(34)
$$
i.e. $\delta_{N}^{0}A_\mu $ coincide with infinitesimal conformal
transformation (3). Symmetries (33) with $n\ge 0$ are in one-to-one
correspondence with the symmetries introduced in Ref.~9. To show this
correspondence in more details, let us discuss the algebraic properties
of the symmetry transformations $\delta_{N}^{n}$.

The commutator of two transformations can be easily calculated
using formula (27). We have
$$
[\delta_{M}^{m}, \delta_{N}^{n}]{\cal F} =\lambda^{-m-n}[\tilde M,
\tilde N]({\cal F}) +m\lambda^{-m-n-1}\tilde N^\lambda \tilde M
({\cal F})- n \lambda^{-m-n-1}\tilde M^\lambda \tilde N ({\cal F}),
\eqno(35)
$$
where $\tilde N^\lambda $ and $\tilde M^\lambda $ are components of
vector fields $\tilde N$ and $\tilde M$ along $\p_\l$. Substituting
(24) into (35), we obtain
$$
[\delta_{M}^{m}, \delta_{N}^{n}]=\delta_{[M, N]}^{m+n}, \
m,n,=0,\pm 1, \pm 2,...
\eqno(36)
$$
for $M,N\in {\cal A}=\{P_\mu , X_a, D\}$, i.e. $\delta_{M}^{m}$
with generators $M$ from the 8-dimensional Lie algebra $\cal A$
(see Ref.~9) form the affine Lie algebra ${\cal A}\otimes C[\l ,\l^{-1}]$.
For generators $\delta_{Y_a}^{m}$ we have
$$
[\delta_{Y_1}^{m}, \delta_{Y_1}^{n}] = i (m-n) (\delta_{Y_1}^{m+n-1}
-\delta_{Y_1}^{m+n+1}),
\eqno(37a)
$$
$$
[\delta_{Y_2}^{m}, \delta_{Y_2}^{n}] = (m-n)
(\delta_{Y_2}^{m+n-1}
+\delta_{Y_2}^{m+n+1}),
\eqno(37b)
$$
$$
[\delta_{Y_3}^{m}, \delta_{Y_3}^{n}] = 2i(m-n) \delta_{Y_3}^{m+n},
\eqno(37c)
$$
$$
[\delta_{Y_1}^{m}, \delta_{Y_2}^{n}] = \delta_{[Y_1, Y_2]}^{m+n}+
m(\delta_{Y_1}^{m+n-1} +\delta_{Y_1}^{m+n+1})-in(\delta_{Y_2}^{m+n-1}
-\delta_{Y_2}^{m+n+1}),
\eqno(38a)
$$
$$
[\delta_{Y_1}^{m}, \delta_{Y_3}^{n}] = \delta_{[Y_1, Y_3]}^{m+n}+
2im\delta_{Y_1}^{m+n} -in(\delta_{Y_3}^{m+n-1}-\delta_{Y_3}^{m+n+1}),
\eqno(38b)
$$
$$
[\delta_{Y_2}^{m}, \delta_{Y_3}^{n}] = \delta_{[Y_2, Y_3]}^{m+n}+
2im\delta_{Y_2}^{m+n} - n(\delta_{Y_3}^{m+n-1}+\delta_{Y_3}^{m+n+1}),
\eqno(38c)
$$
$$
[\delta_{Y_1}^{m}, \delta_{N}^{n}] = \delta_{[Y_1, N]}^{m+n}
- in(\delta_{N}^{m+n-1}-\delta_{N}^{m+n+1}),
\eqno(39a)
$$
$$
[\delta_{Y_2}^{m}, \delta_{N}^{n}] = \delta_{[Y_2, N]}^{m+n}
- n(\delta_{N}^{m+n-1}+\delta_{N}^{m+n+1}),
\eqno(39b)
$$
$$
[\delta_{Y_3}^{m}, \delta_{N}^{n}] = \delta_{[Y_3, N]}^{m+n}
- 2in\delta_{N}^{m+n},
\eqno(39c)
$$
where $N\in {\cal A}=\{P_\mu , X_a, D\}$. Notice that the imaginary
unit $i$ can be removed from structure constants in (37)-(39) by the
following change of generators $\delta_{Y_1}^{m}\rightarrow
i\delta_{Y_1}^{m},\ \delta_{Y_3}^{m} \rightarrow i\delta_{Y_3}^{m}$
(see Ref.~9). Then ``half" of the algebra (36)-(39) with $m,n\ge 0$ will
coincide with the one considered in Ref.~9.  Let us emphasize that in
(36)-(39) the numbers $m,n,=0, \pm 1,...$ are any integer numbers. It
is important to undestand that the change
$\delta_{Y_1}^{m}\rightarrow i\delta_{Y_1}^{m},\ \delta_{Y_3}^{m}
\rightarrow i\delta_{Y_3}^{m}$ is possible only by considering
complex gauge fields or by transition from the signature $(4,0)$ to
the signature $(2,2)$.

{}Finally, for generators $K_\mu$ of special conformal
transformations, the commutators of transformations
$\delta_{K_\mu}^{m}$ with other transformations will be again the
symmetry transformations, but with coefficients depending on the
coordinates on $\cal Z$.  For example,
$$ [\delta_{P_1}^{m},
\delta_{K_1}^{n}]=\delta_{[P_1,K_1]}^{m+n}+
\frac{m}{2}v_{+}^{2}\delta_{P_1}^{m+n+1}
-\frac{m}{2}v_{+}^{1}\delta_{P_1}^{m+n},
\eqno(40)
$$
where $v_{+}^{1}=y-\lambda\bar z, v_{+}^{2}=z+\lambda\bar y$ are
holomorphic coordinates on $U_+\subset \cal Z$. This means that the
structure constants in all the commutators with $\delta_{K_\mu}^{m}$
are replaced by the ``structure functions", that has been noticed by
Popov and Preitschopf.$^{9}$ We do not write out these commutators
looking similar to (40). In twistor terms, commutators like (40)
appear because of the fact that the expansion in Laurent series in
$\lambda$ and formulae like (27), (29) etc. may be used only by the
existence of the holomorphic projection $p: {\cal Z} \rightarrow
CP^1$ leading to the distinguishing of coordinate $\lambda$.  Special
conformal transformations $K_\mu$ and transformations
$\delta_{K_\mu}^{m}$ related to $K_\mu$ do not preserve the bundle
${\cal Z} \rightarrow CP^1$ and make invalid the formulae based on
the expansion in $\lambda$. On formal level, this is connected with
the fact that special conformal transformations transform
$\lambda$ into a function on $x_\mu$ and $\lambda$, and this leads to
appearence of ``structure functions" in formulae like (40). For
correct discussion of symmetries related to $K_\mu$, it is necessary
to pass from the flat hyperK\"ahler space $R^{4,0}$ to the
four-sphere $S^4$ which is conformally flat (but not hyperK\"ahler)
manifold.

Remind that for real gauge fields  with values in the Lie algebra $su(n)$
the transition functions ${\cal F}$ in the bundle $\tilde E$ have to
satisfy the condition (14c). Generally speaking, infinitesimal
transformations do not (infinitesimally) preserve this condition, since
$$
(\delta_{N}^{n}{\cal F}\mid_{\lambda\rightarrow-
\frac{1}{\bar \l}})^\dagger = (-1)^n\l^n\tilde N{\cal F}(\l
)=(-1)^n\delta_{N}^{-n}{\cal F}(\l ).
\eqno(41) $$
However, we can introduce infinitesimal symmetry transformations
$\delta_{N_R}^{n}$ preserving reality conditions by setting
$$
\delta_{N_R}^{n}{\cal F}:=\delta_{N}^{n}{\cal F}+(-1)^n
\delta_{N}^{-n}{\cal F}=(\lambda^{-n}+(-\lambda )^n)\tilde N{\cal F}
\equiv N_{n}^{R}{\cal F}.
\eqno(42)
$$
{}For transformations (42), the formulae for $\phi_{N_{n}^{R}}$,
$\delta_{N_R}^{n}\psi_\pm$ and $\delta_{N_R}^{n}A_\mu$ will have the
same form as the formulae (28)-(33) with the replacement $N_n$ by
$N_{n}^{R}$, and that is why we shall not write out them. As to the
algebraic properties of transformations (42), then, by direct
calculation, it can be shown that generators
$\delta_{P_{\mu}^{R}}^{n}$, $\delta_{X_{a}^{R}}^{n}$ and
$\delta_{D^{R}}^{n}$ form a closed algebra.  Commutation relations
analogous to (36)-(39), which we do not write out, easily follow from
the definition (42).

\vspace{0.6cm}

{\bf V. CONCLUSION}

\vspace{0.3cm}

Symmetries of the SDYM and the self-dual gravity equations are
important in quantization of Yang-Mills model,  N=2 strings and
related models, that has been discussed, for example, in recent
papers.$^{18,19}$  We think that the twistor point of view on symmetries
of the SDYM equations used in this paper clarifies the geometric
and algebraic sence of the so called `hidden symmetries' and makes
them not so mystical. It would be interesting to consider the affine
Lorentz symmetries$^{10}$ of the self-dual gravity equations from the
twistor point of view.  Notice that all the results of our paper may
be generalized on the SDYM equations in the $4k$-dimensional spaces
considered e.g. in Ref.~20.

\vspace{0.6cm}

{\bf ACKNOWLEDGEMENTS}

\vspace{0.3cm}

The author thanks for their kind hospitality the Institut f\"ur
Physik, Humboldt-Universit\"at zu Berlin, where this work was
initiated, and the Fakult\"at f\"ur Physik, Universit\"at Freiburg,
where  part of this work was done. This work was supported by the
DAAD and the Heisenberg-Landau Program.
\newpage
\noindent
$^1\ \ $ M.K.Prasad, A.Sinha and L.-L.Chau Wang, Phys.Lett. {\bf B87},
             237 (1979); \par Phys.Rev.Lett. {\bf 43}, 750 (1979).\\
$^2\ \ $  K.Pohlmeyer, Commun.Math.Phys. {\bf 72}, 37 (1980).\\
$^3\ \ $ L.-L.Chau, M.-L.Ge and Y.-S.Wu, Phys.Rev. {\bf D25},
           1086 (1982);\par
      L.-L.Chau and Y.-S.Wu, Phys.Rev. {\bf D26}, 3541 (1982);\par
      L.-L.Chau, M.-L.Ge, A.Sinha and Y.-S.Wu, Phys.Lett.
      {\bf B121}, 391 (1983);\par
      L.-L.Chau, Lect. Notes Phys. Vol.189, 111 (1983).\\
$^4\ \ $  L.Dolan, Phys.Lett. {\bf B113}, 387 (1982); Phys.Rep.
      {\bf 109}, 3 (1984).\\
$^5\ \ $  K.Ueno and Y.Nakamura, Phys.Lett. {\bf B109}, 273 (1982);\par
         K.Takasaki, Commun.Math.Phys. {\bf 94}, 35 (1984). \\
$^6\ \ $ J.Avan, H.J.de Vega and J.M.Maillet, Phys.Lett. {\bf B171}, 255
        (1986); \par J.Avan and H.J.de Vega, Int.J.Mod.Phys. {\bf A3},
        1273 (1988).\\
$^7\ \ $  C.N.Yang, Phys.Rev.Lett. {\bf 38},  1377 (1977).\\
$^8\ \ $  L.Crane, Commun.Math.Phys. {\bf 110}, 391 (1987).\\
$^9\ \ $  A.D.Popov and C.R.Preitschopf, Phys.Lett. {\bf B374}, 71
            (1996).\\
$^{10}\ $  A.D.Popov, M.Bordemann and H.R\"omer, Phys.Lett. {\bf B385},
             63 (1996).\\
$^{11}\ $  N.M.J.Woodhouse, Class.Quantum Grav. {\bf 2}, 257 (1985).\\
$^{12}\ $  M.F.Atiyah, N.J.Hitchin and I.M.Singer, Proc.R.Soc.Lond.
           {\bf A362}, 425 (1978).\\
$^{13}\ $  A.A.Belavin and V.E.Zakharov, Phys.Lett. {\bf B73}, 53 (1978).\\
$^{14}\ $  R.S.Ward, Phys.Lett. {\bf A61}, 81 (1977).\\
$^{15}\ $  L.J.Mason and N.M.J.Woodhouse, {\it Integrability, Self-Duality
      and Twistor Theory}\par (Clarendon Press, Oxford, 1996).\\
$^{16}\ $  M.F.Atiyah and R.S.Ward, Commun. Math.Phys. {\bf 55}, 117
           (1977).\\
$^{17}\ $  M.Legar\'e and A.D.Popov, Phys. Lett. {\bf A198}, 195 (1995);
           JETP Lett. {\bf 59},  883 (1994).\\
$^{18}\ $ V.P.Nair and J.Schiff, Nucl.Phys. {\bf B371}, 329 (1992);
	  Phys.Lett. {\bf B246}, 423 (1990);
          \par A.Losev, G.Moore, N.Nekrasov and S.Shatashvili, In:
      {\it S-Duality and Mirror Symmetry},\par eds. E.Gava, K.S.Narain
      and C.Vafa (Nucl.Phys. {\bf B46} (Proc.Suppl.), 1996);
      \par S.V.Ketov, hep-th/9606142; E.J.Martinec, hep-th/9608017;
      \par D.Kutasov and E.Martinec, hep-th/9612102.\\
$^{19}\ $ T.Inami, H.Kanno, T.Ueno and C.-S.Xiong, hep-th/9610187;
	\par D.Cangemi, hep-th/ 9610021; A.A.Rosly and K.G.Selivanov,
	hep-th/9611101;\par  V.E.Korepin and T.Oota, hep-th/9608064.\\
$^{20}\ $ R.S.Ward, Nucl.Phys. {\bf B236}, 381 (1984);
     \par E.Corrigan, P.Goddard and A.Kent, Commun.Math.Phys.
     {\bf 100}, 1 (1985);\par A.Galperin, E.Ivanov, V.Ogievetsky and
     E.Sokatchev, Ann.Phys.  {\bf 185}, 1 (1988);
     \par A.D.Popov, Mod.Phys.Lett. {\bf A7}, 2077 (1992); \par
     T.A.Ivanova and A.D.Popov, Theor.Math.Phys. {\bf 94}, 225 (1993).

\end{document}